\documentclass[12pt]{spieman}

\usepackage{amsmath,amsfonts,amssymb}
\usepackage{graphicx}
\usepackage{setspace}
\usepackage{tocloft}
\usepackage{lineno}
\usepackage{siunitx}
\title{Optoelectronic Reservoir Computing with an On-Chip True-Time-Delay Element}

\author[a]{Ahmad Murad}
\author[a]{Shadad Watad}
\author[a,b*]{Alina Karabchevsky}
\affil[a]{School of Electrical and Computer Engineering, Ben-Gurion University of the Negev, Beer-Sheva 8410501, Israel}
\affil[b]{Department of Physics, Lancaster University, LA1 4YB, United Kingdom}

\cftpagenumbersoff{figure}
\cftpagenumbersoff{table} 
\begin{document} 
\maketitle

\begin{abstract}
Compact delay elements remain a central challenge in photonic neuromorphic processors. Here, we demonstrate an optoelectronic delayed-feedback reservoir computer that incorporates a foundry-fabricated silicon nitride (Si$_3$N$_4$) true-time-delay circuit within its feedback loop. Eight cascaded Archimedean spirals provide a \SI{1.76}{\metre} on-chip optical path and a calculated passive group delay of \SI{11.63}{\nano\second}. At a feedback gain of $G=0.0223$, the system classifies sinusoidal and square waveforms without error (word error rate, WER\,$=$\,0 across all cross-validation folds), predicts the Mackey--Glass chaotic series with a best-fold normalized mean-square error (NMSE) of $0.0082$, and performs nine-class Japanese Vowels speaker classification with WER\,$=$\,0.0898. We also introduce a subcarrier phase-encoding method that maps the calculated $0.232\pi$ spiral-to-reference phase contrast onto the measured reservoir states through deliberate aliasing, achieving NMSE\,$=$\,0.0767 and WER\,$=$\,0 without increasing the insertion-loss-limited feedback gain. These results show that on-chip propagation delay in Si$_3$N$_4$ can operate as a functional component of an optoelectronic reservoir computer and motivate lower-loss, more-integrated implementations.
\end{abstract}


{\noindent \footnotesize\textbf{*}Alina Karabchevsky,  \linkable{alinak@bgu.ac.il} }

%%%%%%%%%%%%%%%%%%%%%%%%%%  body  %%%%%%%%%%%%%%%%%%%%%%%%%%

\begin{spacing}{2}
%---------------------------------------------------------------------
\section{Introduction}
\label{sec:intro}
\begin{figure*}[t]
  \centering
  \includegraphics[width=\textwidth]{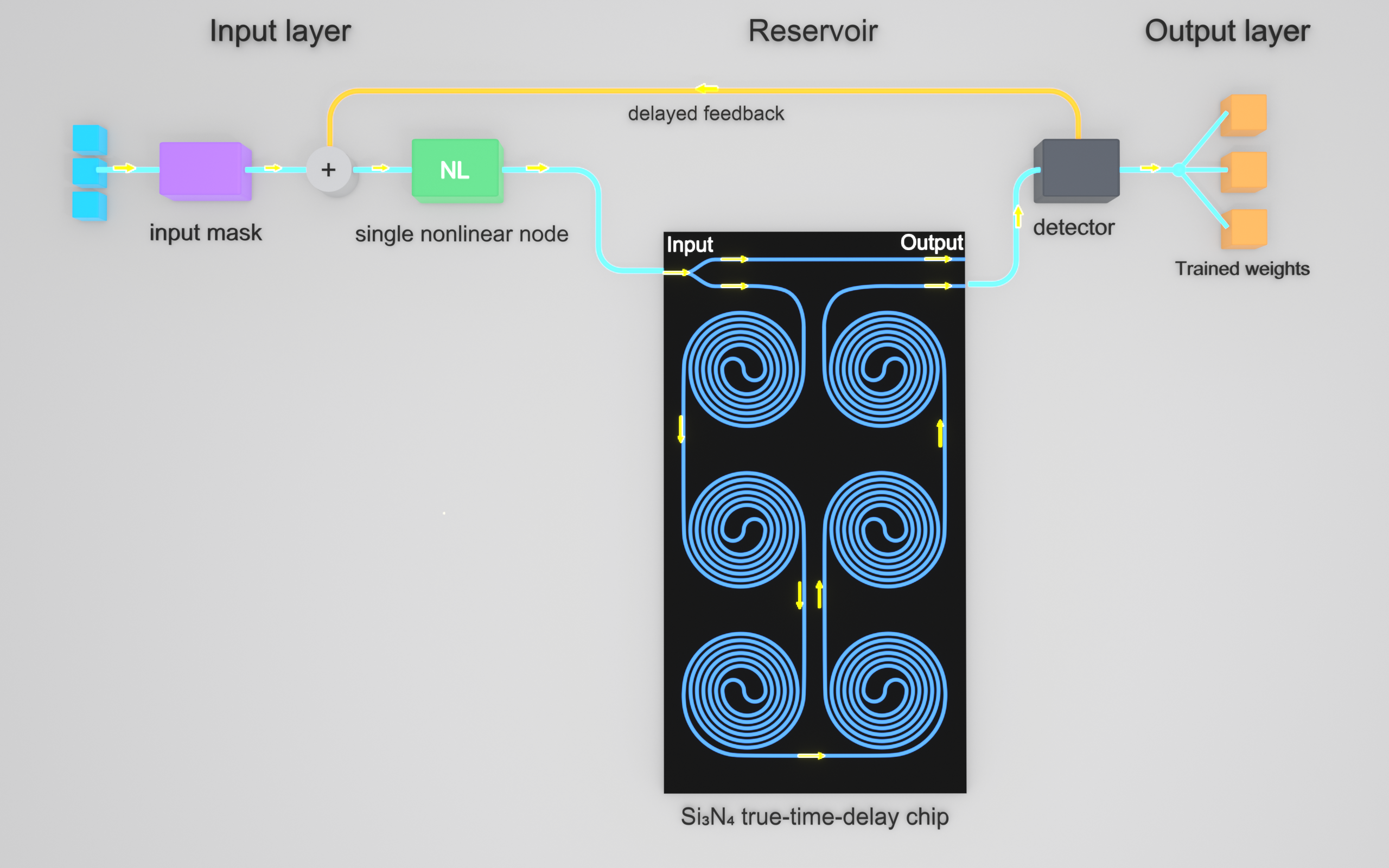}
  \caption{Conceptual illustration of time-multiplexed optoelectronic
  reservoir computing with an on-chip true-time-delay (TTD) element.
  The input $u$ is multiplied by a fixed mask and injected into a single
  nonlinear node, denoted by NL and implemented experimentally using a
  Mach--Zehnder modulator. Delayed feedback generates the virtual-node
  states, which are detected and combined by a trained linear readout to
  produce the output.}
  \label{fig:concept}
\end{figure*}

%---------------------------------------------------------------------
Photonic neuromorphic processing is a candidate technology for
real-time, high-bandwidth, and energy-efficient artificial
intelligence acceleration~\cite{vandoorne2008,appeltant2011,larger2012,vanDerSande2017}.
Among the hardware paradigms proposed, delayed-feedback reservoir
computing (DFRC) reduces the physical hardware to a single nonlinear
node. A feedback delay loop of latency $\tau_D$ expands that node into an
$N$-dimensional virtual processing network by time-multiplexing, and
training is confined to a linear output
layer~\cite{appeltant2011,paquot2012}. The physical delay line supplies
the latency $\tau_D$, and sampling the node output at intervals
$\theta=\tau_D/N$ maps the node dynamics onto $N$ virtual processing
nodes. Figure~\ref{fig:concept} provides a conceptual overview of this
time-multiplexed architecture and highlights the intended role of the
integrated delay element.
 
The feedback delay influences the memory capacity, state
representation, and processing rate of a DFRC
architecture~\cite{duport2012}. Most experimental implementations have
used long optical-fiber spools~\cite{paquot2012,larger2012} or electronic
digital buffers~\cite{morozko2025}. Fiber delays are bulky and cannot be
monolithically integrated, whereas digital buffers require
optical-to-electrical and electrical-to-optical interfaces and do not
preserve optical phase. Incorporating passive on-chip delay elements is
therefore an important step toward compact, high-speed photonic
processors~\cite{mcdonald2020,vanDerSande2017}.
 
Propagation-based true-time delay (TTD) provides a direct mechanism for
processing time-dependent signals on a chip. Because the delay
accumulates through modal propagation, a nonresonant TTD line can provide
an approximately linear phase response over a broad bandwidth and avoid
the resonance-linewidth constraints of slow-light or resonant delay
structures~\cite{melloni2008,ye2015,capmany2023}.
Silicon nitride (Si$_3$N$_4$) on insulator is well suited to meter-scale
photonic routing because of its potential for low
propagation loss, absence of two-photon absorption at telecom
wavelengths, and mature multi-project-wafer (MPW) foundry
ecosystem~\cite{bauters2011,pfeiffer2018,liu2021,xiang2022}. Compact
Archimedean spiral layouts allow meter-scale propagation paths to be
densely integrated into millimeter-scale
footprints~\cite{lee2012,chen2014,murad2025}.
 
Here, we demonstrate an optoelectronic reservoir computer that
incorporates a foundry-fabricated Si$_3$N$_4$ Archimedean-spiral TTD
photonic integrated circuit (LIGENTEC AN350
platform~\cite{ligentec}) within the feedback path of a
Mach--Zehnder-modulator (MZM) node
(Fig.~\ref{fig:concept}). The chip cascades eight spiral waveguides
into a total optical path of $L_{\rm total}\approx\SI{1.76}{\metre}$,
supplying a passive optical group delay
$\tau_D^{\rm optical}=\SI{11.63}{\nano\second}$ at a group index
$n_g=1.981$.
 
We evaluate this platform on four benchmark tasks involving
time-dependent signals. Classification
of sinusoidal and square waveforms is achieved without error, with zero
word error rate on every cross-validation fold. One-step-ahead
prediction of the Mackey--Glass
chaotic series reaches within $8\%$ of the best reported
optoelectronic result. Nine-class Japanese Vowels speaker
classification yields less than half the WER reported for a prior
speckle-based optical processor. These results are obtained at a nonlinear
feedback gain of $G=0.0223$, a factor of more than 40 below the values
at which prior optoelectronic reservoirs operate, showing that useful
computation remains possible well below the strong loop-gain regime
commonly used in previous experiments.
 
We further introduce a subcarrier phase-encoding mechanism specific to
integrated photonic delay. Because a passive waveguide accumulates
true optical phase, a deliberately aliased subcarrier converts the
calculated $0.232\pi$ spiral-versus-reference phase contrast
into reservoir-state information, achieving ${\rm NMSE}=0.0767$ with
zero word error rate without increasing the fixed feedback gain. A
digital delay buffer stores samples but does not itself accumulate
optical propagation phase.
 
To our knowledge, this is the first experimental demonstration of a
foundry-fabricated Si$_3$N$_4$ true-time-delay circuit operating within
the feedback loop of an optoelectronic reservoir computer.

The chip design, fabrication, and optical characterization are detailed
in Sec.~1 of the Supplementary Material.
 
%---------------------------------------------------------------------
\section{Theoretical Framework}
\label{sec:theory}
%---------------------------------------------------------------------
 
\subsection{True-Time Delay in an Integrated Waveguide}
\label{ssec:theory_ttd}
For a guided mode of group velocity $v_g$, the envelope of a modulated
optical signal traversing a waveguide of physical length $L$
accumulates a delay
\begin{equation}
  \label{eq:ttd}
  \Delta t=\frac{L}{v_g}=\frac{n_g L}{c},
  \qquad
  n_g(\lambda)=n_{\rm eff}(\lambda)-\lambda\,\frac{dn_{\rm eff}}{d\lambda},
\end{equation}
where $n_{\rm eff}$ is the modal effective index and $n_g$ the group
index. The delay is governed by the group index of the guided mode
rather than by the modal effective index or the bulk material index of
the nitride. For the cross section used here, the waveguide contributes
a dispersive term $\lambda\,|dn_{\rm eff}/d\lambda|=0.386$, so that
$n_g=1.981$ exceeds both the effective index $n_{\rm eff}=1.595$ and
the bulk nitride index of 1.974 at \SI{1550}{\nano\metre}. Estimating
$\Delta t$ from $n_{\rm eff}$ alone would underestimate it by
approximately $19\%$. Because $n_g$ varies slowly with wavelength for this cross section
(Supplementary Material, Sec.~1.2), the group delay varies across the
modulation bandwidth of the reservoir by less than $10^{-5}$ of its
value. More generally, the delay is set by the physical path length and
the group index, both of which vary slowly with wavelength, so the
usable bandwidth is limited by the material and waveguide dispersion
rather than by a resonance linewidth. Resonant and slow-light delay
elements obtain their delay from a steep dispersion feature and are
therefore restricted to a narrow, thermally sensitive band around the
resonance~\cite{melloni2008,notomi2001}.
 
\subsection{Delayed-Feedback Reservoir Computing}
\label{ssec:theory_dfrc}
In a delayed-feedback optoelectronic reservoir, the state $x(t)$ of a
single nonlinear node obeys the delay-differential
dynamics~\cite{larger2012,paquot2012}
\begin{equation}
  \label{eq:rc_continuous}
  T_R\,\dot{x}(t) + x(t)
  = f\!\bigl[\,G\,x(t-\tau_D) + \rho\,J(t)\,\bigr],
\end{equation}
where $T_R$ is the node response time, $G$ the feedback gain, $\rho$
the input scaling, $\tau_D$ the total loop delay, and
$J(t)$ the masked input. The single physical node is expanded into a
high-dimensional reservoir by time-multiplexing~\cite{appeltant2011}.
Each input sample $u_k$ is held for one delay period and multiplied by
a fixed random binary mask $\mathbf{m}\in\{-1,+1\}^{N}$, producing the
injection signal
\begin{equation}
  \label{eq:mask}
  J(t)=u_k\,m_i,
  \qquad
  (i-1)\theta\;\le\;t-k\tau_D\;<\;i\theta,
\end{equation}
with virtual-node index $i=1,\dots,N$ and node spacing
\begin{equation}
  \label{eq:theta}
  \theta=\frac{\tau_D}{N}.
\end{equation}
Sampling the node output at intervals $\theta$ yields the state vector
$\mathbf{x}_k=[x_k^{(1)},\dots,x_k^{(N)}]^\top$, and stacking these
over $K$ input samples gives the reservoir state matrix
$\mathbf{X}\in\mathbb{R}^{K\times N}$ from which the linear readout is
trained.
 
Equations~(\ref{eq:mask}) and (\ref{eq:theta}) make the role of the
physical delay explicit. The loop delay $\tau_D$ sets the reservoir
dimension $N$ at fixed node spacing, and simultaneously fixes the
temporal window over which the reservoir integrates its own past
output. Information injected at sample $k$ re-enters the node one
delay period later, so $\tau_D$ is the characteristic timescale of the
system's fading memory~\cite{appeltant2011,duport2012}. For
$T_R\ll\tau_D$, as holds for the MZM node used here,
Eq.~(\ref{eq:rc_continuous}) reduces to the discrete map
$x_{n+1}=f[\,Gx_n+\rho u_n\,]$ evaluated once per virtual
node~\cite{paquot2012}.
 
Two properties of the delay element follow directly from this
description, and both are satisfied by construction by the chip of
Sec.~\ref{ssec:chip}. The first is phase linearity. Any frequency
dependence of $\Delta t$ across the modulation bandwidth distorts the
temporal spacing of the virtual nodes and degrades the state
representation. The slowly varying group index established in
Sec.~\ref{ssec:theory_ttd} minimizes delay distortion across the
reservoir passband, whereas resonant delay elements can introduce
group-delay variation near resonance. The second property is low insertion
loss. The loop gain scales linearly with the optical power returned
to the photodetector, so attenuation in the delay element translates
directly into reduced $G$. This coupling motivates the loss-per-delay
figure of merit used throughout this work,
\begin{equation}
  \label{eq:fom}
  \mathrm{FOM}=\frac{\mathcal{L}_{\rm excess}}{\tau_D^{\rm optical}}
  \quad[\si{\decibel/\nano\second}],
\end{equation}
which evaluates to \SI{2.0}{\decibel/\nano\second} for the
present device (Supplementary Material, Sec.~1.5). A useful reservoir
delay line must therefore provide sufficient delay while maintaining
low loss. The central experimental question in this work is whether
useful computational performance can be retained when this trade-off
places the system well below the conventional gain regime.
 
\subsection{Nonlinear Node: MZM Transfer Function and Reservoir Map}
\label{ssec:theory_mzm}
The nonlinear mapping is provided by an MZM operated in
intensity-modulation mode. For a modulator driven by voltage $V_b$, the
transmitted power is~\cite{saleh2019}
\begin{equation}
  \label{eq:mzm_transfer}
  P_{\rm out}(V_b) = \frac{P_{\rm max}}{2}
  \left[1+M\sin\!\left(\frac{\pi}{V_\pi}V_b+\phi\right)\right],
\end{equation}
where $P_{\rm max}$ is the peak power reaching the photodetector,
$M\in(0,1]$ is the extinction parameter, $V_\pi$ is the half-wave
voltage, and $\phi$ is the static phase offset. Substituting the delayed
photodetector voltage as the RF drive yields the reservoir
map~\cite{paquot2012,larger2012}
\begin{equation}
  \label{eq:reservoir_map}
  x_{n+1} = \frac{G}{2}\left[1+M\sin(\pi x_n+\phi_0)\right],
\end{equation}
with nonlinear gain
\begin{equation}
  \label{eq:gain}
  G=\frac{G^{*}P_{\rm max}}{V_\pi},
  \qquad
  G^{*}=\SI{10}{\volt\per\milli\watt},
\end{equation}
where $G$ is the primary control parameter and $G^{*}$ is the combined
transimpedance and voltage gain of the detection and feedback chain.
Iterating Eq.~(\ref{eq:reservoir_map}) at the calibrated values of $M$
and $\phi_0$ gives a single stable fixed point below
$G\approx0.98$, a first period-doubling bifurcation at that value, and
a positive Lyapunov exponent beyond
$G\approx1.27$~\cite{larger2012}. Prior optoelectronic reservoirs have
operated near $G\approx1$~\cite{paquot2012,duport2012}, a regime
selected to maximize nonlinear state mixing. The platform reported here
operates well below that range; the benchmark results in
Sec.~\ref{sec:results} quantify the performance retained in this
low-gain regime.

%---------------------------------------------------------------------
\subsection{Integrated Delay Element}
\label{ssec:chip}
%---------------------------------------------------------------------
The delay element is a foundry-fabricated Si$_3$N$_4$ photonic
integrated circuit realized on the LIGENTEC AN350 MPW
platform~\cite{ligentec}. Eight identical Archimedean spirals, each of
routed length $\approx\SI{0.22}{\metre}$, are cascaded in series to
realize a total on-chip optical path
$L_{\rm total}\approx\SI{1.76}{\metre}$ within the reticle. The
waveguide cross section (\SI{1}{\micro\metre} core width,
\SI{350}{\nano\metre} guiding-layer thickness) supports the fundamental
mode used at $\lambda=\SI{1550}{\nano\metre}$, with
$n_{\rm eff}=1.595$ and a
group index $n_g=1.981$ that varies slowly across the C-band. By
Eq.~(\ref{eq:ttd}) this gives the design delay
\begin{equation}
  \label{eq:tau_main}
  \tau_D^{\rm optical}=\frac{n_g L_{\rm total}}{c}
  =\SI{11.63}{\nano\second}
  \qquad(\SI{1.45}{\nano\second}\ \text{per spiral}).
\end{equation}
A Y-branch splitter at the input divides the signal between the spiral
cascade and a co-fabricated straight reference waveguide
($L_{\rm ref}=\SI{5}{\milli\metre}$). Because both paths share the
input facet, coupling optics, and detection chain, the reference
isolates the spiral excess loss and simultaneously provides the second
optical path exploited by the subcarrier phase-encoding mechanism of
Sec.~\ref{ssec:subcarrier}.
 
Optical spectrum analysis of the two paths under identical coupling and
amplification conditions yields a spiral excess loss of
$\mathcal{L}_{\rm excess}\approx\SI{23.2}{\decibel}$ for this
first-generation, density-optimized layout, corresponding by
Eq.~(\ref{eq:fom}) to \SI{2.0}{\decibel} per nanosecond of
delay. The complete loss budget, layout, post-fabrication inspection,
transmission spectra, and a process-migration pathway toward lower-loss
operation are reported in the Supplementary
Material, Sec.~1. For the reservoir experiments that follow, the
relevant consequence is that this loss sets the optical power at the
photodetector and therefore the loop gain $G$
(Sec.~\ref{sec:setup}). The benchmark results obtained at this fixed
gain are reported in Sec.~\ref{sec:results}.
 
A direct phase-comparison measurement of the differential delay between
the spiral and reference arms produced a nanosecond-scale estimate of
the expected order of magnitude. However, because the spiral-arm signal
amplitude after
\SI{23.2}{\decibel} of attenuation falls below the residual coherent
electrical pickup, that measurement is limited by pickup and does not
support a quantitative determination of $n_g$ (Supplementary Material,
Sec.~2). An in situ consistency check is instead provided by the
reservoir experiment. The \SI{10}{\mega\hertz} subcarrier phase contrast
between the spiral and reference paths (Sec.~\ref{ssec:subcarrier}) is
consistent with the
$\Delta\varphi\approx0.2319\pi$ predicted by
Eq.~(\ref{eq:tau_main}), and is exploited there as a computational
resource.
 
%---------------------------------------------------------------------
\section{Experimental Setup}
\label{sec:setup}
\begin{figure*}[htbp]
  \centering
  \includegraphics[width=\textwidth]{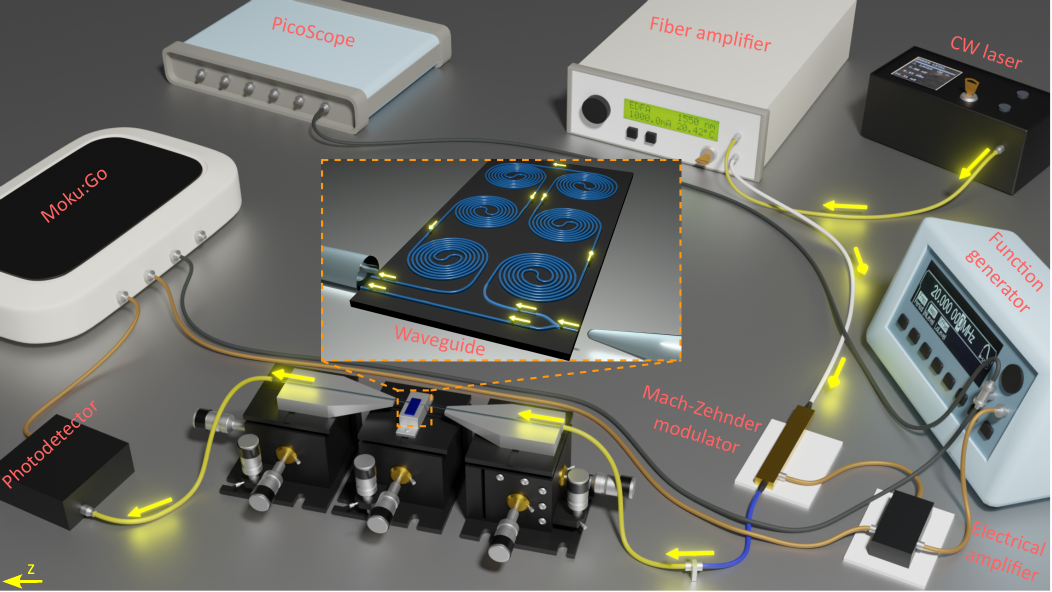}
  \caption{Conceptual three-dimensional rendering of the experimental
  optoelectronic reservoir computer; components and optical paths are
  not drawn to scale. The amplified continuous-wave laser light
  (EDFA, top) is modulated by the Mach--Zehnder modulator and coupled
  into the Si$_3$N$_4$ true-time-delay chip via lensed single-mode
  fibers mounted on three-axis translation stages (center). The inset
  shows the chip with its cascaded Archimedean spiral waveguides. Six
  spirals are drawn for visual clarity, whereas the fabricated chip
  contains eight cascaded spirals. The transmitted light is detected by a
  fiber-coupled photodetector, and the electronic feedback loop is
  closed through the acquisition instrument, the arbitrary waveform
  generator, and the summing amplifier driving the MZM RF port
  (Sec.~\ref{sec:setup}). Arrows indicate the direction of optical
  propagation.}
  \label{fig:setup}
\end{figure*}

%---------------------------------------------------------------------
Figure~\ref{fig:setup} illustrates the optoelectronic reservoir computer with
the Si$_3$N$_4$ TTD chip in the optical feedback path. A
continuous-wave laser at $\lambda\approx\SI{1553}{\nano\metre}$ is
amplified by an erbium-doped fiber amplifier (EDFA,
$\approx\SI{30}{\milli\watt}$ maximum output) and passed through a
fiber polarization controller into the optical input of the
Mach--Zehnder modulator (Thorlabs LN81S-FC,
$V_\pi\approx\SI{5.37}{\volt}$, \SI{10}{\giga\hertz} bandwidth), which
serves as the nonlinear node of the reservoir
(Sec.~\ref{ssec:theory_mzm}). The modulated light is coupled into the
chip through a lensed single-mode fiber aligned to the input edge
coupler by a three-axis translation stage, propagates through the
cascaded spiral delay line
($\tau_D^{\rm optical}=\SI{11.63}{\nano\second}$,
Sec.~\ref{ssec:chip}), and is collected at the output edge coupler by a
second lensed fiber feeding a photodetector (Thorlabs DET08CFC,
$10^3$\,V/A transimpedance gain, \SI{5}{\giga\hertz} bandwidth).
 
The electrical section closes the feedback loop and injects the input
data. The photodetector output is acquired by a Moku:Go instrument
(Liquid Instruments), whose finite-impulse-response (FIR) filter implements
the electronic portion of the feedback delay and whose auxiliary
outputs supply the MZM DC bias (setting $\phi_0$) and the
variable-attenuator control. Its filter and acquisition parameters are
treated as a fixed, calibrated stage of the loop. The delayed feedback
signal is summed in a passive resistive network with the masked input
waveform $J(t)$ generated by an arbitrary waveform generator (Agilent
33220A) in triggered burst mode, and the sum is amplified by an LM7171
op-amp driver ($\pm\SI{14.7}{\volt}$ supply) before driving the MZM RF
port. A PicoScope PS5000a supplies the \SI{5}{\hertz} burst-trigger
clock that synchronizes the waveform generator and the acquisition,
acting as the experiment's timing master. The total feedback delay therefore comprises the electronically realized portion together with the chip's passive optical delay. At the best validated operating point (\(N=50\) virtual nodes, node spacing \(\theta=20.48~\mathrm{ns}\)), the chip delay corresponds to \(\tau_D^{\mathrm{optical}}/\theta=0.57\) of one virtual-node period. Thus, although the optical delay is shorter than one complete node interval, it represents a significant fraction of the virtual-node spacing and cannot be neglected on the timescale of the reservoir dynamics. The chip therefore contributes to the temporal relation between consecutive reservoir states and, consequently, to the reservoir dynamics and memory.

With the chip in the loop, the optical power reaching the photodetector
is $P_{\rm max}\approx\SI{0.012}{\milli\watt}$, placing the reservoir
at a nonlinear gain of $G=0.0223$ (Eq.~(\ref{eq:gain})), lower by more
than a factor of 40 compared to the $G\approx1$ regime in which prior
optoelectronic reservoirs have been
operated~\cite{paquot2012,duport2012,morozko2025}. This provides a
stringent test of the architecture, since every benchmark in
Sec.~\ref{sec:results} is obtained under a feedback gain far weaker
than those commonly reported for previous implementations. Calibration
parameters $\{P_{\rm max},V_\pi,M,\phi\}$ were extracted by recording
$P_{\rm out}$ versus $V_b$ with the chip in the optical path and
fitting Eq.~(\ref{eq:mzm_transfer}) by nonlinear least squares
(Supplementary Material, Sec.~3). Only
the linear output layer is trained by ridge regression
$\mathbf{W}_{\rm out}=(\mathbf{X}^\top\mathbf{X}+\lambda\mathbf{I})^{-1}
\mathbf{X}^\top\mathbf{y}$~\cite{lukosevicius2009}. Two metrics are
reported. The normalized mean-square error
${\rm NMSE}=\langle(y-\hat y)^2\rangle/{\rm Var}(y)$~\cite{paquot2012}
measures how closely the continuous readout tracks the target, and the
word error rate (WER) counts the fraction of misclassified instances.
The two are not redundant, since a readout can fall on the correct side
of the decision threshold on every sample while remaining numerically
inaccurate, and WER is quoted only for the classification tasks.
The hyperparameters $\{\rho,\phi_0,\lambda\}$ and, where relevant, the
delay-length setting, were optimized by Tree-structured Parzen
Estimator (TPE) search~\cite{bergstra2011} directly on hardware, and
all results were evaluated by cross-validation.

\section{Results}
\label{sec:results}
Here, we present the results of in situ optimization of the
experimental reservoir-computing system incorporating the integrated
Si$_3$N$_4$ true-time-delay chip across four benchmark tasks. The
experimentally measured performance is compared with matched numerical
simulations where applicable. Owing to the insertion loss introduced by
the integrated delay line, the nonlinear feedback gain was fixed at
$G=0.0223$. For each benchmark, the phase bias $\phi_0$, input scaling
$\rho$, and readout regularization parameter $\lambda$ were optimized
directly on the experimental system using the Tree-structured Parzen
Estimator Bayesian optimization algorithm.

\subsection{Waveform Classification}
\label{ssec:sinsquare}

\begin{figure}[htbp]
    \centering
    \includegraphics[width=\textwidth]{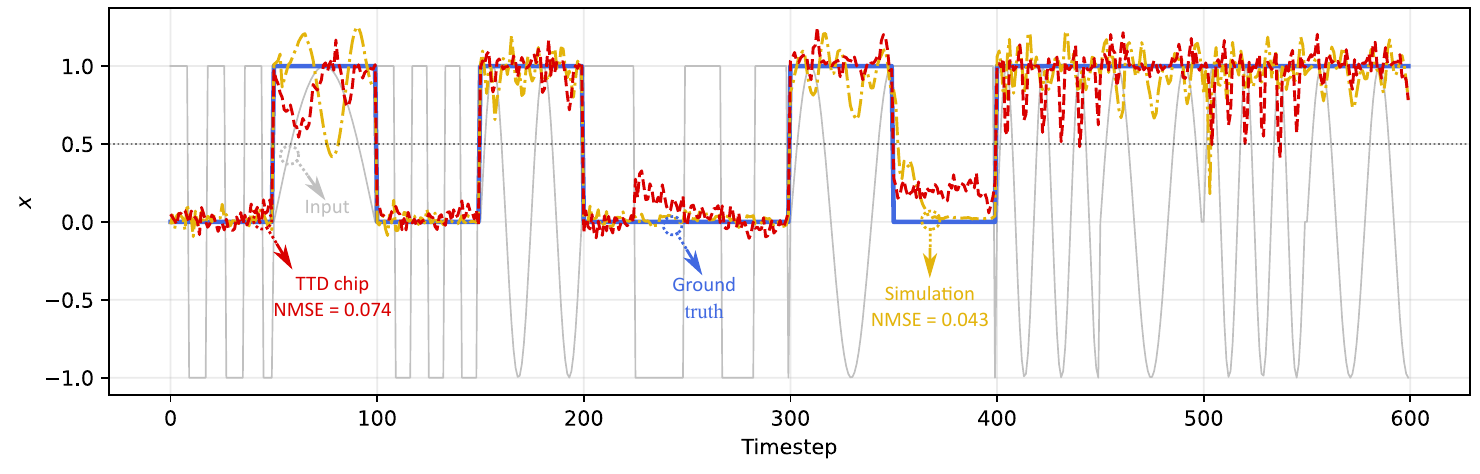}
    \caption{Best-fold experimental readout for sinusoidal and square-wave classification using the integrated TTD reservoir. The target assigns $y=1$ to sinusoidal waveforms and $y=0$ to square waveforms. The readout achieved an NMSE of $0.074$ and a WER of $0$. The dotted horizontal line marks the decision threshold at $y=0.5$.}
    \label{fig:ttd_sinsquare_bestfold}
\end{figure}

The first benchmark evaluated the discrimination between sinusoidal and
square waveforms at varying frequencies, following
Refs.~\cite{paquot2012,morozko2025}. The dataset comprised 24
waveforms, including 12 sinusoidal and 12 square-wave signals, each
represented by $N=50$ virtual-node samples and evaluated using
five-fold cross-validation. At the validated TTD-chip operating point
of $G=0.0223$, the reservoir correctly classified all test waveforms
across the five folds, yielding $\mathrm{WER}=0$. The best-performing
fold achieved an NMSE of $0.074$, while the mean NMSE across all folds
was $0.0913$. As shown in Fig.~\ref{fig:ttd_sinsquare_bestfold}, the
experimental readout follows the transitions between the two waveform
classes and remains on the correct side of the decision threshold, with
only small fluctuations around the binary target levels.

\subsection{Subcarrier Phase Encoding}
\label{ssec:subcarrier}

\begin{figure*}[htbp]
    \centering
    \includegraphics[width=\textwidth]{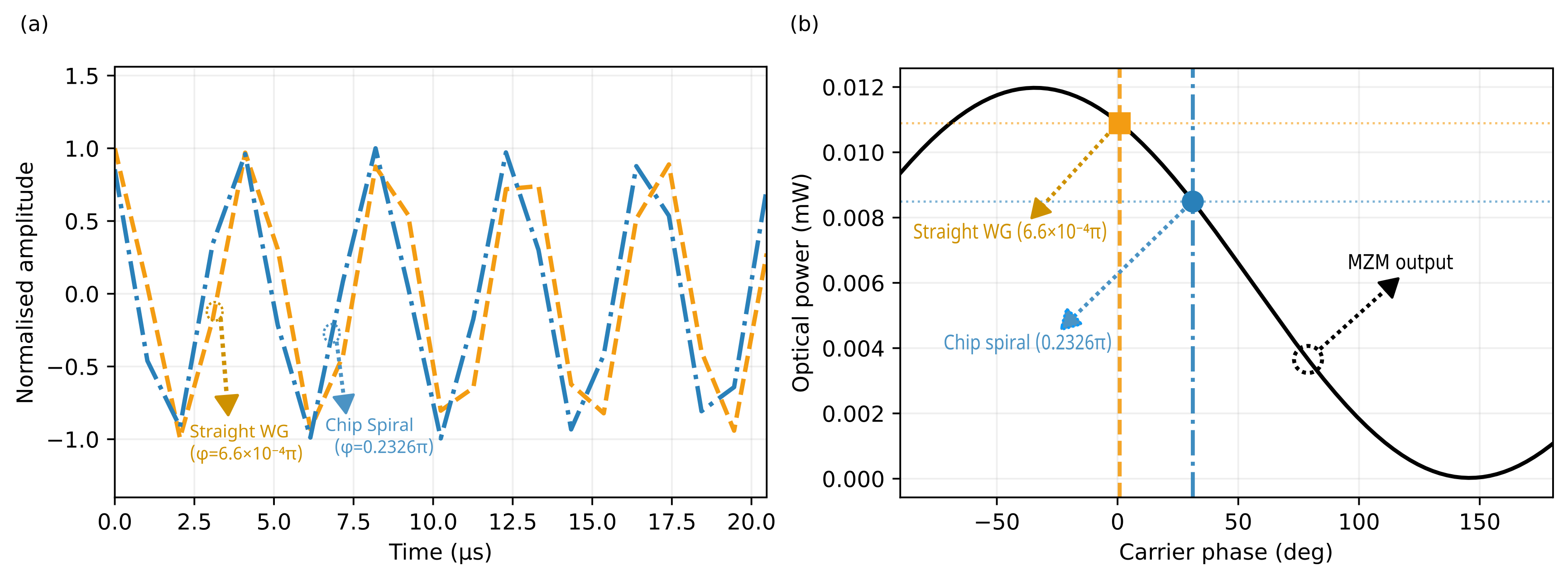}
    \caption{Subcarrier phase-encoding mechanism. (a) A \SI{10}{\mega\hertz} carrier sampled at \SI{976.6}{\kilo\hertz} aliases to \SI{234}{\kilo\hertz}. The calculated delays of the spiral and straight-reference paths correspond to phase shifts of $0.2326\pi$ and $6.6\times10^{-4}\pi$, respectively, and a phase contrast of $\Delta\varphi=0.2319\pi$. (b) The mean MZM output as a function of carrier phase illustrates how this path-dependent phase relation maps onto distinguishable output levels in the measured reservoir states.}
    \label{fig:ttd_subcarrier_mechanism}
\end{figure*}

The second benchmark investigated whether the path-dependent phase of
the integrated TTD chip could provide an additional encoding mechanism.
A \SI{10}{\mega\hertz} sinusoidal carrier was superimposed on the
reservoir input mask. From the calculated propagation delays, the
spiral and the co-fabricated \SI{5}{\milli\metre} straight-waveguide
reference accumulate phase shifts of $0.2326\pi$ and
$6.6\times10^{-4}\pi$, respectively, giving a predicted phase contrast
of $\Delta\varphi=0.2319\pi$. At the acquisition sampling rate of
\SI{976.6}{\kilo\hertz}, the \SI{10}{\mega\hertz} carrier was aliased
to \SI{234}{\kilo\hertz} while preserving the path-dependent phase relation between
the two optical paths. As illustrated in
Fig.~\ref{fig:ttd_subcarrier_mechanism}, the MZM transfer function
converts this phase relation into distinguishable output states,
allowing the chip's physical path response to be encoded without
increasing the insertion-loss-limited feedback gain.

Using the same sinusoidal and square-wave dataset and a
50-trial TPE optimization performed directly on the experimental
system, the phase-encoding configuration achieved a best-fold NMSE of
$0.0767$ and a mean NMSE of $0.0992$, while maintaining
$\mathrm{WER}=0$ across all folds. During the optimization, the mean
NMSE decreased from an initial random-probe value of $0.654$ to
$0.0992$, a reduction by a factor of $6.6$. These results show that the
path-dependent phase relation provides a measurable encoding channel at
the fixed operating gain of $G=0.0223$.

\subsection{Japanese Vowels Speaker Classification}
\label{ssec:jv}

The third benchmark evaluated the integrated TTD reservoir on the
Japanese Vowels speaker-classification task. The dataset comprised 640
variable-length sequences of 12 linear-prediction cepstral
coefficients, recorded from nine speakers pronouncing the vowel /ae/,
with each sequence classified according to speaker
identity~\cite{kudo2000}. The hyperparameters were optimized using a
50-trial TPE search at the calibrated operating point of $G=0.0223$,
with each candidate evaluated as the mean of three experimental
repetitions. A matched noise-free simulation was subsequently performed
using the same optimized operating parameters.

\begin{figure}[H]
    \centering
    \includegraphics[width=\textwidth]{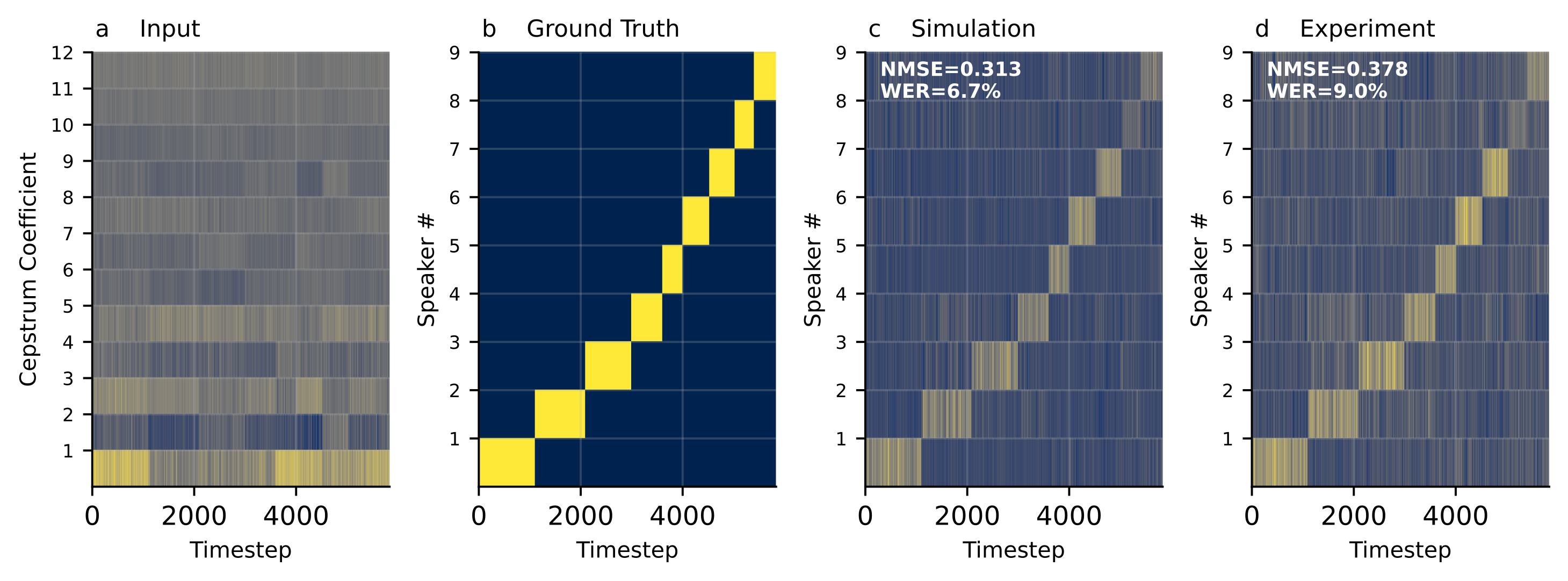}
    \caption{Japanese Vowels speaker classification for the test set,
    sorted by speaker identity for visual clarity. (a) Input sequences
    of 12 linear-prediction cepstral coefficients. (b) Target speaker
    identities represented by one-hot encoding. (c) Experimental
    TTD-chip readout (WER $=0.0898$). (d) Matched noise-free simulation
    (WER $=0.0670$). Panels (b)--(d) share the same color scale. Correct
    classifications follow the diagonal pattern in the target matrix,
    whereas errors appear as off-diagonal responses.}
    \label{fig:ttd_jv_waveforms}
\end{figure}

The experimental TTD reservoir achieved a WER of $0.0898$ and an NMSE
of $0.3778$, whereas the matched numerical simulation yielded a WER of
$0.0670$ and an NMSE of $0.3127$. As shown in
Fig.~\ref{fig:ttd_jv_waveforms}, both readouts reproduce the overall
speaker-classification pattern. The experimental output contains
additional off-diagonal responses associated with misclassifications.
The approximately $2.3$-percentage-point difference between the
experimental and simulated WER is consistent with contributions from
measurement noise and hardware imperfections at the same
insertion-loss-limited operating point.

For comparison, the previously reported optimal-gain implementation
achieved a WER of $0.0650$ and an NMSE of $0.2710$ at
$G\approx0.52$~\cite{morozko2025}, using an FPGA-based digital delay
line in place of an optical one. Despite operating at a nonlinear gain
lower by a factor of 23, the present experimental system remains within
$2.5$ percentage points of that reference WER and outperforms the WER
of $0.1850$ reported for a speckle-based optical
implementation~\cite{paudel2020}. Moreover, the matched simulation
differs from the optimal-gain reference by only $0.2$ percentage points
in WER, suggesting that much of the additional experimental
classification error originates from hardware noise and measurement
imperfections rather than from the low-gain operating regime alone.

\subsection{Mackey--Glass Time-Series Prediction}
\label{ssec:mg}

As the final benchmark, the integrated TTD reservoir was evaluated for
one-step-ahead prediction of the Mackey--Glass chaotic time series,
generated according to

\begin{equation}
    \dot{x}(t) =\frac{\beta_{\mathrm{MG}}x\!\left(t-\tau_{\mathrm{MG}}\right)}{1+\left[x\!\left(t-\tau_{\mathrm{MG}}\right)\right]^n}-\gamma x(t)
    \label{eq:mackey_glass}
\end{equation}

where $\beta_{\mathrm{MG}}=0.2$, $\gamma=0.1$, $n=10$, and
$\tau_{\mathrm{MG}}=17$~\cite{mackey1977}. At each time step, the
current value $x(t)$ was supplied as the reservoir input, while the
subsequent value $x(t+1)$ served as the prediction target. Consecutive
samples of the generated series differ by only a few percent, so the
target is strongly correlated with the input and the benchmark measures
how faithfully the reservoir reproduces a smooth deterministic
trajectory. Generalization performance was evaluated by
cross-validation, with the task-specific hyperparameters optimized
directly on the experimental system using TPE.

\begin{figure*}[htbp]
    \centering
    \includegraphics[width=\textwidth]{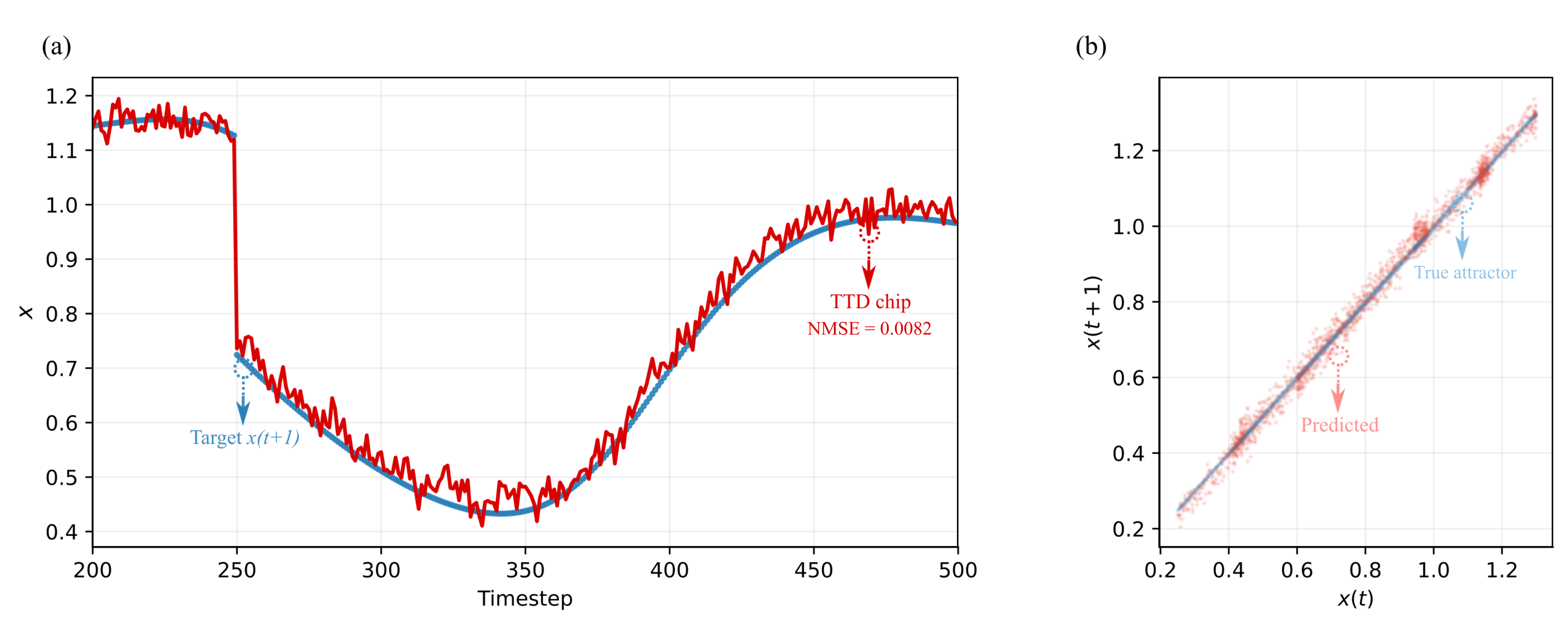}
    \caption{Best-fold one-step-ahead prediction of the Mackey--Glass chaotic time series using the integrated TTD reservoir. (a) Representative 300-time-step window showing the target $x(t+1)$ and the experimental reservoir prediction (NMSE $=0.0082$). (b) Phase-space plot of $x(t+1)$ versus $x(t)$ for the target and predicted trajectories. The two distributions overlap closely over the range of the series.}
    \label{fig:ttd_mg_bestfold}
\end{figure*}

At the validated TTD-chip operating point of $G=0.0223$, the reservoir
achieved a best-fold NMSE of $0.0082$ and a mean NMSE of $0.0112$. This
represents the lowest NMSE obtained among the experimental benchmarks
considered in this work. As shown in
Fig.~\ref{fig:ttd_mg_bestfold}, the best-fold prediction closely
follows the target sequence across both slowly varying regions
and sharper turning points, with the remaining deviations occurring
predominantly as small, high-frequency fluctuations rather than
systematic temporal lag or amplitude mismatch.

The corresponding phase-space plot of $x(t+1)$ against $x(t)$ shows the
predicted and true points occupying the same narrow band about the
diagonal, which is the form taken by both series at this sampling
interval. The panel shows that the readout reproduces the range and
distribution of the target without evident bias or amplitude
compression over the full excursion of the series. For comparison,
Paquot \textit{et al.} reported an NMSE of $0.0076$ at
$G\approx1$~\cite{paquot2012}. The present best-fold result is
therefore within approximately $8\%$ of that value despite operation at
the substantially lower gain of $G=0.0223$, demonstrating accurate
one-step-ahead prediction in the insertion-loss-limited operating
regime.

\section{Discussion and Conclusion}
\label{sec:discussion}
The benchmark results show that an integrated Si$_3$N$_4$ photonic TTD
line can operate within the feedback loop of a reservoir computer and
retain useful computational performance despite substantial insertion
loss. Waveform classification is achieved without classification
errors, and the best-fold Mackey--Glass prediction error is within
approximately $8\%$ of the reported fiber-spool reference. On the
Japanese Vowels speaker-classification task, the present architecture
achieves less than half the WER reported for a speckle-based optical
processor and remains within 2.5 percentage points of a digital-delay
implementation operating at a nonlinear gain higher by a factor of 23.

The subcarrier phase-encoding mechanism exposes an intrinsic advantage
of physical optical delay structures. By directly exploiting the native
optical phase accumulated across the integrated waveguide, this
mechanism provides a phase-sensitive encoding channel in the physical
domain. A digital electronic memory buffer stores samples but does not
itself accumulate optical propagation phase. The channel remains usable
at the fixed gain of $G=0.0223$ and does not require an increase in the
feedback gain.

In conclusion, we have demonstrated an optoelectronic reservoir
computer incorporating a foundry-fabricated Si$_3$N$_4$ integrated TTD
circuit within its feedback loop. The results show that a compact,
passive propagation delay can contribute to time-multiplexed reservoir
dynamics at a feedback gain far below the conventional regime. They
also identify a route toward more highly integrated photonic reservoir
architectures. The principal engineering objective for a
second-generation device is to reduce the \SI{23.2}{\decibel} excess
loss, which sets the operating gain
and limits the metrology accessible on the present chip.

\subsection*{Disclosures}
The authors declare no competing interests.

\subsection*{Code and Data Availability}
The code and data that support the findings of this study are not
publicly available at this time but may be obtained from the
corresponding author upon reasonable request.

\subsection*{Author Contributions}
A.M. and S.W. contributed equally to this work, sharing the methodology,
investigation, and formal analysis. All authors wrote the original draft
of the manuscript. A.K. conceptualized and supervised the project, acquired funding, reviewed and edited the manuscript.

\subsection*{Acknowledgments}
The authors thank Dr. Aviad Katiyi for rendering the graphical illustrations used in this manuscript.

\subsection*{Funding}
The research was supported by the Israel Science Foundation ISF (1023/24).

%%%%%%%%%%%%%%%%%%%%%%% References %%%%%%%%%%%%%%%%%%%%%%%%%

\end{spacing}
\end{document}